\documentstyle[aps,epsf,multicol]{revtex}
\begin{document}
\renewcommand{\thefootnote}{\fnsymbol{footnote}}
\draft
\title{
Origin of the Charge-Orbital Stripe Structure in La$_{1-x}$Ca$_{x}$MnO$_{3}$
($x=1/2, 2/3$)
}
\author{
Tetsuya Mutou $^{1}$ and Hiroshi Kontani $^{2}$
}
\address{
$^{1}$ The Institute of Physical and Chemical Research (RIKEN),
Wako, Saitama 351-0198, Japan \\
$^{2}$ Institute for Solid State Physics, University of Tokyo,
7-22-1 Roppongi, Minato-ku, Tokyo 106-8666, Japan \\
}
\date{\today}

\maketitle

\begin{abstract}
We propose the origin of the charge-ordered stripe structure
with the orbital ordering observed experimentally 
in La$_{1-x}$Ca$_{x}$MnO$_{3}$ 
($x=1/2, 2/3$), in which the long-range Coulomb interaction
plays an essential role. 
We study a Hubbard model with doubly-degenerate $e_{g}$ orbitals,
and treat the on-site Coulomb interaction ($U$) and 
the nearest-neighbor one ($V$) with the Hartree-Fock 
approximation.
Both the charge and orbital ordering structures observed 
in experiments are reproduced in a wide region of the $U$-$V$ phase 
diagram determined by the present study. 
The stability of the orbital ordering is also confirmed by
the perturbation theory.
\end{abstract}

\pacs{
PACS numbers: 71.20.Be, 71.10.-w, 71.10.Fd
}

\begin{multicols}{2}
\narrowtext

In some perovskite-type hole-doped manganese oxides 
$R_{1-x}A_{x}$MnO$_{3}$ ($R$ : rare earth elements, $A$ : 
alkaline earth elements), the colossal magnetoresistance 
effect has been the subject of intense studies.
Recently, characteristic charge ordering phenomena 
in these materials have also attracted a growing interest.
Especially, in La$_{1-x}$Ca$_{x}$MnO$_{3}$ for $x=1/2$ and $2/3$,
it has been reported that the charge-orbital stripe (COS) structure 
occurs in some periodicities which correspond to the commensurate
concentrations
\cite{Chen-PRL96,Radaelli97,Ramirez96,Chen-JAppl97,Mori-Nature98,Fernandez99}.

Concerning the pure LaMnO$_{3}$ system ($x=0$), 
the antiferromagnetic (AF) insulating phase appears \cite{Wollan55},
where the orbital ordering accompanied with the Jahn-Teller (JT) 
distortion coexists because of the JT active ion Mn$^{+3}$
\cite{LaMn03,stabilizedJT}. 
In the same way, it is suggested
that the JT effect is also the origin of the stripe 
structure in the systems with finite carrier concentrations
such as $x=1/2$ and $2/3$ \cite{Mori-Nature98,Hotta98}.  
According to this scenario, however, 
a considerably strong JT effect is necessary to realize 
the insulating COS
structure \cite{Hotta98}. 
In this sense, 
it is insufficient to ascribe the origin of the stripe structure
only to the JT effect.

\begin{figure}
\epsfxsize=60mm
\centerline{\epsffile{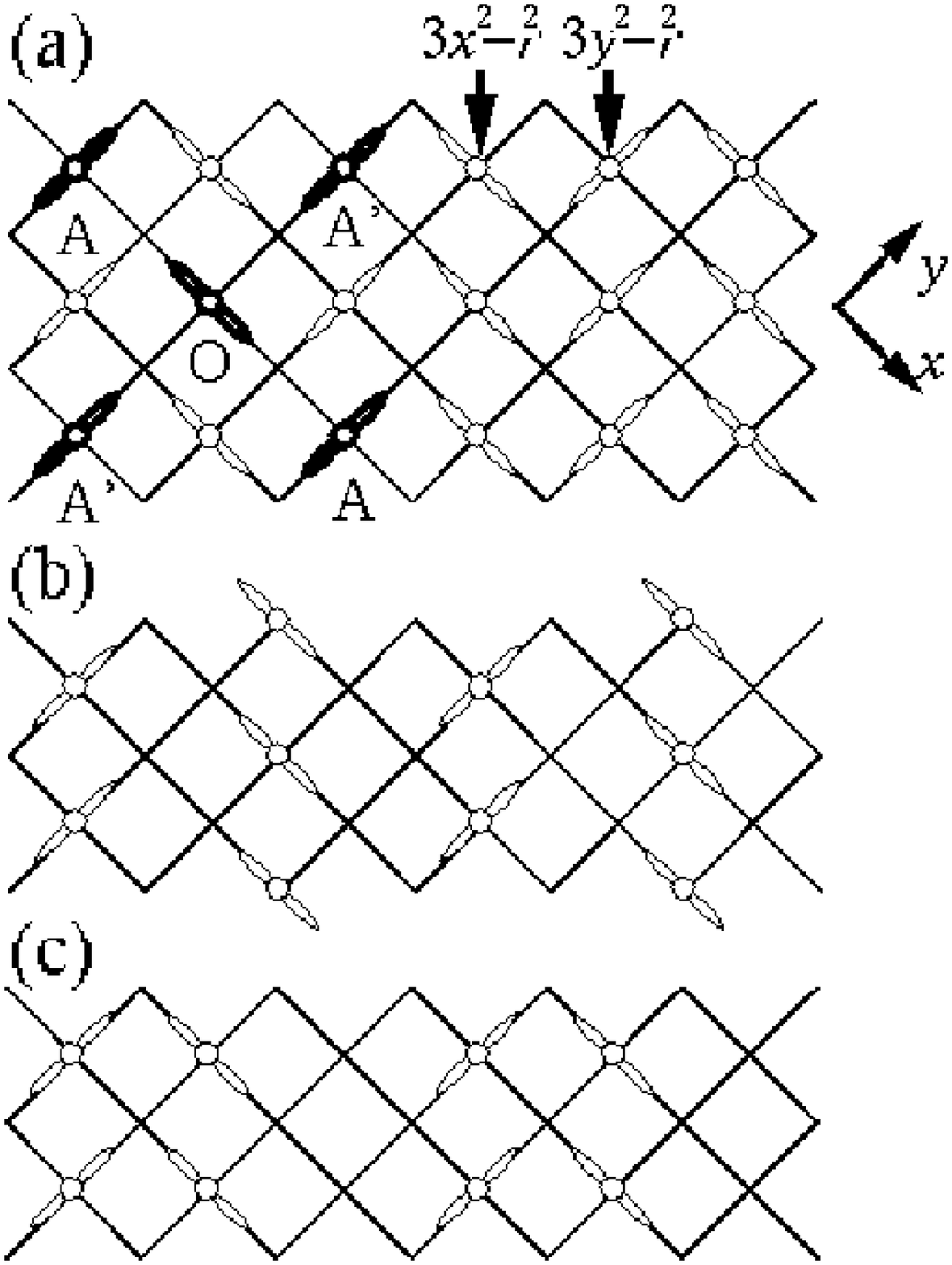}}
\caption{
(a) Schematic configuration of the COS structure 
observed in experiments for (a) the $x=1/2$ system and (b) the 
$x=2/3$ system. 
(c) Paired COS structure reported in 
Ref.[5] for the $x=2/3$ system.
}
\label{fig:COpattern}
\end{figure}

In the present letter, we show that 
the COS structure observed in the La$_{1-x}$Ca$_{x}$MnO$_{3}$ 
can be explained only by considering 
the Coulomb interaction between carriers.
We study a Hubbard model with doubly degenerated orbitals which
correspond to $e_{g}$ orbitals on Mn ions.
By treating the Coulomb interaction
with the Hartree-Fock approximation,
we investigate the stability of the COS structure 
observed in experiments
for the $x=1/2$ system (Fig.\ \ref{fig:COpattern}(a))
\cite{Chen-PRL96,Radaelli97,Comment} and the $x=2/3$ system 
(Fig.\ \ref{fig:COpattern}(b)) \cite{Fernandez99,Comment}, and
discuss the another type of the charge ordering
found by electron microscopy study 
reported in Ref.\ \cite{Mori-Nature98} 
(Fig.\ \ref{fig:COpattern}(c)).
We show that the COS structure appears 
for the realistic strength of on-site and nearest-neighbor 
Coulomb interactions.
In particular, we emphasize that 
the nearest-neighbor Coulomb interaction is indispensable
for the occurrence of the stripe structure with the orbital ordering.

Several authors have studied 
the effect of the on-site Coulomb interaction
on the orbital ordering in manganese oxides previously
\cite{Mizokawa95,Ishihara,Koshibae97,Maezono98}.
Especially, in the $x=0$ system, 
it was shown that the orbital ordering is
stabilized by the on-site Coulomb interaction
\cite{Koshibae97}. However, the effect of the 
long-range Coulomb interaction have not been studied enough.

The charge ordering in La$_{1-x}$Ca$_{x}$MnO$_{3}$ systems
is the ordering of Mn ions with different valences: Mn$^{+3}$ and
Mn$^{+4}$. 
Three electrons in $t_{2g}$ orbitals 
construct the localized $S=3/2$ spin.
The strong Hund coupling works between the localized $t_{2g}$ 
spin and the $S=1/2$ spin of the itinerant $e_{g}$ electron.
The difference of valences between Mn$^{+3}$ and Mn$^{+4}$
corresponds to whether the Mn ion has an $e_{g}$ electron or not.

In the COS phase observed in both $x=1/2$ and $2/3$ systems,
the COS structures are formed in all the $a$-$b$ planes, and 
they are stacked along the $c$-axis without misfitting. 
Thus, for simplicity, we investigate the charge
configuration in the two-dimensional system which corresponds 
to the $a$-$b$ plane. The charge ordering is also observed in 
layered-type manganese oxides \cite{Moritomo95}.

\begin{figure}
\epsfxsize=60mm
\centerline{\epsffile{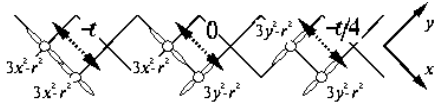}}
\caption{
Hopping integrals between nearest-neighboring sites.
}
\label{fig:Ham}
\end{figure}

The Hamiltonian which we consider is expressed as follows;
\begin{eqnarray}
  \label{eqn:FullHam}
  \cal{H} &=& \sum_{\stackrel{\scriptstyle \langle i,j \rangle}
    {(\alpha,\beta)=(X,Y)}}
  (t^{\alpha \beta}_{ij} c^{\dag}_{i \alpha}c_{j \beta} + 
  {\rm H.c.}) \nonumber \\
  &+& U\sum_{i}n_{i X}n_{i Y}
  + V\sum_{\langle i,j \rangle}(n_{i X}+n_{i Y})(n_{j X}+n_{j Y}),
\end{eqnarray}
where $n_{i \alpha}$ denotes a number operator
$c^{\dag}_{i \alpha}c_{i \alpha}$ on the site $i$.
Indices $\alpha$ and $\beta$ correspond to two orbitals of $e_{g}$ 
: $3x^{2}-r^{2}$ and $3y^{2}-r^{2}$ 
symbolized by $X$ and $Y$, respectively. 
In the first term of the Hamiltonian,
$t^{\alpha \beta}_{ij}$ is 
the hopping integral between the orbital
$\alpha$ on the site $i$ and 
$\beta$ on $j$.
We define hopping integrals between nearest-neighbor sites
as follows;
\begin{eqnarray}
  \label{eqn:hoppingXX}
  t^{XX}_{ij} &=& \left\{
  \begin{array}{ll}
    -t & (\mbox{\boldmath$R$}_{j}=\mbox{\boldmath$R$}_{i} \pm 
    \hat{\mbox{\boldmath$x$}}) \\
    -t' & (\mbox{\boldmath$R$}_{j}=\mbox{\boldmath$R$}_{i}
    \pm \hat{\mbox{\boldmath$y$}}),
  \end{array}
  \right. \\
  \label{eqn:hoppingXY}
  t^{XY}_{ij} &=& 0, \\
  \label{eqn:hoppingYY}
  t^{YY}_{ij} &=& \left\{
  \begin{array}{ll}
    -t' & (\mbox{\boldmath$R$}_{j}=\mbox{\boldmath$R$}_{i} \pm 
    \hat{\mbox{\boldmath$x$}}) \\
    -t & (\mbox{\boldmath$R$}_{j}=\mbox{\boldmath$R$}_{i}
    \pm \hat{\mbox{\boldmath$y$}}),
  \end{array}
  \right. 
\end{eqnarray}
where $\hat{\mbox{\boldmath$x$}}$ and $\hat{\mbox{\boldmath$y$}}$
denote unit vectors of $x$ and $y$ directions, respectively
(Fig.\ \ref{fig:Ham}).
We introduce values of $t$ and $t'$ as $t=t_{0}$ and $t'=t_{0}/4$
\cite{Ishihara97}.
Then, the band width $W$ of the free system ($U=V=0$)
is equal to $5t_{0}$. Hereafter we take $t_{0}$ as the energy unit. 

In the insulating phase, the screening effect of 
the Coulomb interaction is expected to be suppressed.
Thus, not only the on-site Coulomb 
interaction $U$ but also the nearest-neighbor one $V$
becomes important.
We assume for simplicity 
that the nearest-neighbor Coulomb interaction is not 
depend on orbitals. 

The Hamiltonian (\ref{eqn:FullHam}) is based on the 
double exchange model \cite{DE}. For the strong Hund coupling 
$J_{\rm H}$ ($\sim 1 {\rm eV}$),
it is expected that the spin of the electron in the
$e_{g}$ band is fixed to be parallel to the local $t_{2g}$ spin
at $T \ll J_{\rm H}$. Thus we neglect the spin degeneracy
for both the ferromagnetic and the paramagnetic states. 
In other words, we consider the spinless fermion system \cite{Zang96}.
The present model can be interpreted as a single-band Hubbard model
with spin-dependent hopping integrals provided the orbital
degree of freedom is expressed as the pseudo-spin.
Below, we show that the above simplified model can reproduce
the COS structure observed in manganese oxides.

The carrier concentration $n$ is given by 
\begin{equation}
  n=\frac{1}{2}\sum_{\alpha=X,Y}\frac{1}{N}\sum_{i=1}^{N}
  \langle n_{i \alpha} \rangle,
  \label{eqn:meanvalue}
\end{equation}
where $N$ denotes the number of sites.
In this letter, we treat two cases with carrier concentrations
$n=1/4$ and $1/6$ which correspond to systems with $x=1/2$ and $2/3$
in La$_{1-x}$Ca$_{x}$MnO$_{3}$, respectively.
Hereafter, 
we apply the Hartree-Fock approximation to the two terms with
Coulomb interactions $U$ and $V$ 
in the Hamiltonian (\ref{eqn:FullHam}), 
and determine $U$-$V$ phase diagrams.

\begin{figure}
\epsfxsize=60mm
\centerline{\epsffile{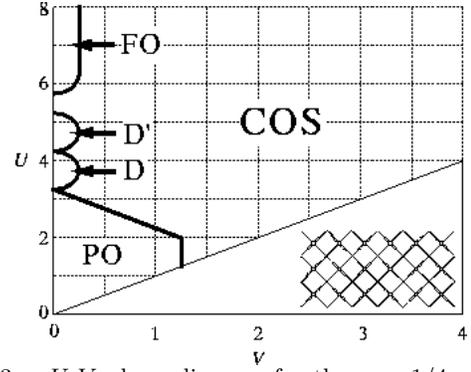}}
\caption{
$U$-$V$ phase diagram for the $n=1/4$ case. The COS phase
corresponds to the COS phase with the 
orbital ordering. Only the FO and PO phases are metallic.
The inset shows 
the schematic charge configuration in the COS phase
(see also Fig.1(a)).
}
\label{fig:PD-4}
\end{figure}

First we show the result for the $n=1/4$ case which corresponds
to the La$_{1/2}$Ca$_{1/2}$MnO$_{3}$ system.
In the phase diagram Fig.\ \ref{fig:PD-4}, we see the COS phase, 
whose structure is schematically displayed
in Fig.\ \ref{fig:COpattern}(a) or the inset of Fig.\ \ref{fig:PD-4},
spreads out in the wide region. 
The COS phase is realized generally in 
many $x=1/2$ compounds of $R_{1/2}A_{1/2}$MnO$_{3}$
(see Fig.\ \ref{fig:COpattern}(a)), 
and it is called the CE type \cite{Wollan55} 
except for the spin configuration.

There is the ferro-orbital (FO) phase 
around $V=0$ for $U \gtrsim 6$, where 
the orbital ordering is realized 
but the charge ordering does not occur; $n_{i}^{X(Y)} \simeq 1$
and $n_{i}^{Y(X)} \simeq 0$ for all sites.
We comment that this FO phase, which appears only in the region
$U \gtrsim W (=5)$, may be the artifact of the mean-field 
approximation.
In reality, in a single-band Hubbard model, 
ferromagnetism is hardly realized beyond the mean-field 
approximation \cite{ferro}. 
On the other hand, the COS phase is realized for
$U < W$, which means the validity of the COS phase 
beyond the mean-field approximation.

For smaller $U$, the para-orbital (PO) phase is
realized, where neither
charge ordering nor orbital ordering exist :
$n_{i}^{X} = n_{i}^{Y} = n {\rm (: const.)}$ for all sites.
For the limited region of $U \sim W$ and the small $V$, 
D and D' phases exist. 
We do not mention their structures here.
Note that FO and PO phases are metallic.

\begin{figure}
\epsfxsize=60mm
\centerline{\epsffile{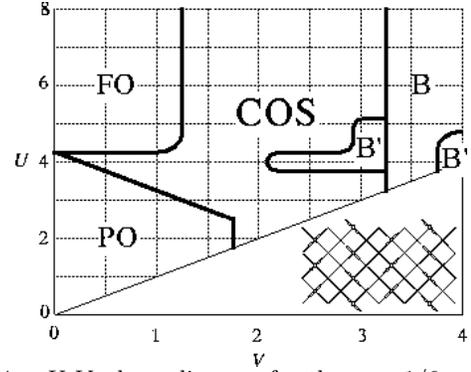}}
\caption{
$U$-$V$ phase diagram for the $n=1/6$ case. 
Only the FO and PO phases are metallic.
The inset shows the schematic charge configuration in the COS phase
for the $n=1/6$ case (see also Fig.1(b)).
}
\label{fig:PD-6}
\end{figure}

Let us turn to the case with the carrier concentration $n=1/6$.
In the recent neutron diffraction experiment for 
La$_{1/3}$Ca$_{2/3}$MnO$_{3}$ \cite{Fernandez99}, 
the stripe structure corresponding Fig.\ \ref{fig:COpattern}(b)
is realized undoubtedly.
In the present result, the same COS structure is reproduced 
in the fairly wide region of the $U$-$V$ diagram as shown in
Fig.\ \ref{fig:PD-6}.
We emphasize that the region of the COS phase includes
the realistic values of $U (\sim 5)$ and $V$, and 
it is expected to spread further 
if we include the long-range Coulomb interaction beyond
the nearest-neighbor one.

For $V \sim U$, another types of the charge 
ordering denoted by B and B' occur. 
However we do not mention these structures 
since they appear only for unrealistically larger
values of $V$.
Similarly to the $n=1/4$ case, 
there is the PO phase with neither charge nor 
orbital orderings for smaller values of $U$ in the $n=1/6$ case.
The FO phase also exists
for $U \gtrsim W$ and $V \gtrsim 1$. In the $n=1/6$ case,
however, the regions for these phases spread out more
widely than those for the $n=1/4$ case. 

The mechanism of the COS structure for $n=1/4$ displayed 
in Fig.\ \ref{fig:COpattern}(a) can be understood by
the perturbation treatment as follows.
We assume $t \ll U,V$ and set for simplicity $t'=0$
in eq.\ (\ref{eqn:hoppingYY}).
For large values of $U$ and $V$, it is natural that
carriers order alternatively on lattice sites and each 
orbital on a site is singly occupied. In the configuration
shown in Fig.\ \ref{fig:COpattern}(a), the ground-state energy
is given by 
$E_{g}=-2t^{2}/(3V)-[2/(9UV^{2})+29/(45V^{3})+4/\{9V^{2}(U+4V)\}]t^{4}
+{\cal O}(t^{6})$.
If the $Y$-orbital is occupied 
on the O-site (Fig.\ \ref{fig:COpattern}(a)) instead of $X$, 
the ground state energy
is raised by 
$\Delta E_{g} = \{4/(9UV^{2})+8/(45V^{3})\}t^{4}
+{\cal O}(t^{6})$,
since the energy-gain through the exchange processes
with the $Y$-orbital electrons at (A,A')-sites are reduced.
For the orbital ordering of the $n=1/6$ case, 
a similar argument also 
holds in the order of $t^{6}$ provided the charge ordering 
is constructed. 
Thus we understand the reason why the long-range 
Coulomb interaction stabilizes the COS structure with the 
orbital ordering.

Let us discuss another pattern of the COS structure
shown in Fig.\ \ref{fig:COpattern}(c),
which is realized on the surface of the sample in 
La$_{1/3}$Ca$_{2/3}$CuO$_{3}$ according to Ref.\ \cite{Mori-Nature98}. 
This paired COS does not appear in the present phase diagram 
(Fig.\ \ref{fig:PD-6}) in which the JT distortion is neglected.
Concerning the JT distortion,
each non-paired Mn$^{3+}$ stripe causes significant lattice
distortion in its immediate neighborhood and
the overall strain energy will be lowered by forming periodic
array of pairs on Mn$^{3+}$ stripes separated by undistorted 
regions of Mn$^{4+}$ ions \cite{Mori-Nature98}.
We note that the JT distortion is stronger than 
the bulk one, since the elastic constant 
on the surface is smaller.

\begin{figure}
\epsfxsize=60mm
\centerline{\epsffile{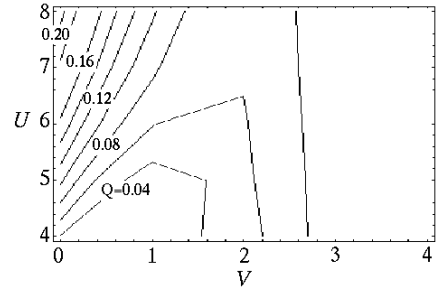}}
\caption{
Contour diagram for values of $Q$ which are needed to realize
the paired COS structure for ($U$, $V$). For detail, see the text.
}
\label{fig:contour}
\end{figure}

In order to estimate the strength 
of the JT distortion energy enough to construct 
the paired stripe structure,
we consider the following JT Hamiltonian ${\cal H}_{\rm JT}$.
\begin{equation}
  {\cal H}_{\rm JT} = Q\sum_{\langle \langle i,j \rangle \rangle}
  \{(n_{iX}-n_{iY})-(n_{jX}-n_{jY})\},
  \label{eqn:HamJT}
\end{equation}
where $\langle \langle \cdots \rangle \rangle$ denotes the sites
which correspond to the paired stripe shown in 
Fig.\ \ref{fig:COpattern}(c); 
in a pair $\langle \langle i,j \rangle \rangle$, sites $i$ and $j$ are 
the left and right site of the pair, respectively. 
In the Hamiltonian (\ref{eqn:HamJT}),
$Q$ denotes the potential energy which corresponds to the 
electron-lattice coupling related to the linear displacement 
of the JT distortion \cite{Millis96}.
For simplicity, we neglect the mixing term between 
two orbitals $3x^{2}-r^{2}$ and 
$3y^{2}-r^{2}$ to realize the configuration
in Fig.\ \ref{fig:COpattern}(c).
The lattice elastic energy,
which is proportional to the square of the displacement,
is neglected.
For the system added the above Hamiltonian (\ref{eqn:HamJT}),
we calculate the energy under the constraint;
$n_{iX}=n_{jY}$ and 
$n_{iY}=n_{jX}$ for $\langle \langle i,j \rangle \rangle$.
Figure\ \ref{fig:contour} shows the contour diagram for values of $Q$
which are needed to realize the paired stripe structure for certain values
of ($U$,$V$) with the same ground-state energy of the 
$Q=0$ system without the constraint.

Obviously, in the case of $V \ll U$, fairly large values of $Q$ are 
needed to realize the paired stripe structure.
On the other hand, for the region of the COS phase 
in Fig.\ \ref{fig:PD-6},
the paired stripe structure can be realized with
relatively small values of $Q$. 
This indicates that 
the nearest-neighbor Coulomb interaction is also important to
construct the paired stripe structure.
The result suggests 
the possibility that the paired stripe structure reported in
Ref.\ \cite{Mori-Nature98} on the surface is realized 
owing to the smaller elastic constant on the surface than
that in the bulk.

Finally, we explain some experiments on the COS phase
and discuss the validity of the long-range Coulomb interaction
scenario;
in Pr$_{1-x}$Ca$_{x}$MnO$_{3}$ systems,
it was reported that
the charge-ordering transition temperature ($T_{\rm CO}$)
is suppressed with external pressure \cite{MoritomoPRB55}.
Under pressure, the transfer integral is expected to become large,
and both $U/t_{0}$ and $V/t_{0}$ decrease. 
In fact, absolute values of the resistivity are suppressed
with increasing pressure \cite{MoritomoPRB55}. 
The reason why the charge ordering becomes unstable is expected
to be the reduction of $U/t_{0}$ and $V/t_{0}$ in the phase
diagram, Figs.\ \ref{fig:PD-4} and \ref{fig:PD-6}.
The present conclusion is qualitatively consistent with the above
experimental result.
The similar behavior is also reported in layered-type 
manganese oxides with chemical pressure \cite{MoritomoPRB56}.

On the other hand, 
recently, the charge-ordering phenomenon was observed in thin
films (thicknesses are $500 \sim 2000$ \AA) in
Nd$_{0.5}$Sr$_{0.5}$MnO$_{3}$ \cite{Prellier99}. 
It was shown
that $T_{\rm CO}$
almost does not change when the thickness of the sample changes,
although the JT distortion varies sensitively with 
the thickness of the sample. 
This is consistent with the present result
because we show that 
the COS structure can appear 
without taking account of the JT effect.
It is suggested that the JT effect is not so important
to the construction of the COS structure observed in manganese oxides
with $x \ge 0.5$.

We would like to comment on that we treated the spinless system
in the present study.
In manganese oxides in which the charge-ordered phase is
observed, $T_{\rm CO}$ is higher than the AF transition temperature 
for $x > 0.5$ \cite{Ramirez96}, 
which will mean that the AF ordering
is a phenomenon in the lower energy scale.
It is of course expected that the RKKY interaction, which is neglected
in the present study, becomes effective to the AF ordering 
at lower temperatures.

In summary, we have shown that the COS phase appears 
in a wide region of the $U$-$V$ phase diagram for both $n=1/4$ and 
$1/6$ cases.
We conclude that the COS structure observed 
in La$_{x}$Ca$_{1-x}$MnO$_{3}$ for $x=1/2$ and $2/3$ is originated 
from the Coulomb interaction.
It is emphasized that doubly-degenerated $e_{g}$ orbitals and
the nearest-neighbor Coulomb interaction are important to realize 
the COS structure.
According to the present scenario,
the COS structure is realized by the long-range
Coulomb interaction at first. 
After that, the JT distortion occurs so that 
the COS structure is stabilized.
In conclusion, the JT distortion is the {\it consequence}
of the COS structure but not the {\it origin} of it. 

We are grateful to K. Ueda for valuable comments.
One of the author (T.M.) is supported by the Special Postdoctoral
Researchers Program from RIKEN.
A part of the numerical calculations was performed on the
super-computer VPP700E of RIKEN.

\end{multicols}

\begin{thebibliography}{99}
\bibitem{stabilizedJT} According to recent theoretical studies, 
the orbital ordering for the $x=0$ system is stabilized
by the JT distortion \cite{Mizokawa95,Ishihara,Koshibae97}.
\bibitem{Comment} From results shown in 
Ref.\ \cite{Radaelli97,Fernandez99}, it is supposed that
the orbital ordering accompanied with the charge ordering
schematically displayed in Fig.\ \ref{fig:COpattern}(a) and (b)
occurs in both $x=1/2$ and $2/3$ systems.
\end{thebibliography}
\end{document}